\documentclass[pra,twocolumn,showpacs,groupedaddress,superscriptaddress,aps,10pt]{revtex4-1}
\usepackage{bm,graphicx,amsmath}
\usepackage{placeins}
\usepackage{amssymb}
\usepackage{amsmath}
\usepackage{epsfig}
\usepackage{amssymb}
\usepackage{color}
\usepackage[colorlinks,linkcolor=blue,citecolor=blue]{hyperref}
\usepackage{subfigure}

\begin{document}

\title{Optical vault: reconfigurable bottle beam by conically refracted light}
\date{\today}

\author{A. Turpin}
\affiliation{Departament de F\'isica, Universitat Aut\`onoma de Barcelona, Bellaterra, E-08193, Spain}
\author{V. Shvedov}
\affiliation{Laser Physics Center, Research School of Physics and Engineering, Australian National University, Canberra ACT 0200, Australia}
\author{C. Hnatovsky}
\affiliation{Laser Physics Center, Research School of Physics and Engineering, Australian National University, Canberra ACT 0200, Australia}
\author{Yu. V. Loiko}
\affiliation{Departament de F\'isica, Universitat Aut\`onoma de Barcelona, Bellaterra, E-08193, Spain}
\author{J. Mompart}
\affiliation{Departament de F\'isica, Universitat Aut\`onoma de Barcelona, Bellaterra, E-08193, Spain}
\author{W. Krolikowski}
\affiliation{Laser Physics Center, Research School of Physics and Engineering, Australian National University, Canberra ACT 0200, Australia}

\begin{abstract} 
We employ  conical refraction of light in a biaxial crystal to create an
optical bottle for  trapping and manipulation of  particles.
We show that by just varying the   polarization of the input light the  bottle can be  opened and closed at will. We experimentally demonstrate stable photophoretic trapping and controllable loading  and unloading of light absorbing  particles in the trap.\\
\textbf{ocis}: (350.4855) Other areas of optics : Optical tweezers or optical manipulation, (050.1940) Diffraction and gratings : Diffraction, (140.7010) Lasers and laser optics : Laser trapping,
(140.3300) Lasers and laser optics : Laser beam shaping, (260.1440) Physical Optics : Birefringence .
\end{abstract}

\date{\today}
\maketitle

\section{Introduction}

Since its inception in the late 70s the field of  optical trapping and manipulation of micron and submicron-sized objects with light has experienced an intense interest and rapid development~\cite{Ashkin:97,Grier:nature:03}. Optical tweezers utilising the presence of mechanical forces arising  from light interaction with matter are now an indispensable tool in various  physical, biological and medical applications. They have been extensively used  in manipulating colloidal particles, molecules ,nanoparticles and even  single atoms. The recent decade has seen an enormous progress in the field of trapping resulting in  the implementation of advanced techniques involving for instance, multiple holographic traps, optical fibers, or singular scalar and vector beams~\cite{Dholakia:08,McGloin:opn:10, Woerdemann:lpr:13}. Optimum conditions for particle trapping are dictated by the optical properties of the particles and the surrounding medium, as well as the physical nature of the light-mediated trapping forces. For instance, while high light intensity attracts and traps transparent high-index objects in a low-index medium, it in fact repels low index particles in a high index environment~\cite{Grover:08}. That is why hollow (or donut) beams are used for efficient trapping in the latter case. In general, depending on the particular media and application, robust trapping and manipulation of micro-objects requires tailoring the light beam intensity pattern via phase and amplitude modulation or by varying spatial coherence of light~\cite{Cizmar:np:10,McGloin:opex:03,Woerdemann:lpr:13,Rao:oc:08,Shvedov:opex:10}.

In 2000, Arlt and Padgett \cite{Arlt:ol:00} introduced the concept of an optical bottle which  represents an optical beam with a low (ideally null) intensity region surrounded entirely by light. Such a bottle could be used as a three-dimensional trap. Following this idea various practical implementations of optical bottles have been proposed. The low intensity regions have been formed using, for instance, interference of multiple laser beams, partially spatially coherent optical vortices or laser beams affected by optical aberrations. The suitability of an optical bottle for particle trapping and manipulation has  been confirmed in experiments with  atoms~\cite{Olson:pra:07,Xu:ol:10,Li:ol:12} and  absorbing particles~\cite{Shvedov:opex:11}. The concept of an optical bottle has also been recently extended to  plasmonic structures~\cite{Genevet:opex:13}.

The problem with an ideal bottle beam is that the more efficiently it traps particles the more difficult it is to actually load it with the particles. Once an optical bottle is formed it actually prevents  particles from entering it ~\cite{McGloin1:far:10}. To cope with this issue, one straightforward solution consists of turning on the bottle beam when the particles already float in the region where the trap will be formed. Another and much more convenient choice would be to design a bottle in such a way that it could be partially opened and closed so it could be loaded and unloaded with particles as required.

The purpose of this work is to prove that such a design is indeed possible. We  will demonstrate that an optical bottle formed  by using conically refracted light in a biaxial crystal could be controlled so it can be opened and closed at will and in real time by varying the polarization of the input beam.  We will then use photophoretic trapping to demonstrate loading and unloading of airborne particles into and from the bottle.

\section{Theory}

Conical refraction refers to the phenomenon associated with the propagation of a collimated optical beam along one of the optical axes of a biaxial crystal~\cite{Berry:pio:07}. Conical refraction was first studied by Hamilton, who predicted that light emerging from the crystal should exhibit conical-like intensity distribution, which was observed soon afterwards by Lloyd [J. G. O'Hara, "The prediction and discovery of conical refraction by William Rowan Hamilton and Humphrey Lloyd," Proc. R. Ir. Acad. 82 (2), 231–257 (1982).]. The geometric optic approximation for the conical refraction ring radius is given by $R_0 = l \alpha$, being $l$  the length of the biaxial crystal and $\alpha$ its conicity. 
A close observation of the ring pattern reveals that it is indeed formed by two concentric bright rings, the so-called fine Poggendorff splitting. 
In fact, for a focused input beam, the bright ring with the Poggendorff splitting appears at the focal plane of the lens which is a symmetry plane of the beam evolution. Out of this plane, two bright on-axis spots are observed symmetrically at both sides which are named as Raman spots. 
It turns out that the actual light intensity distribution outside the crystal is strongly determined by the initial parameters of the beam such as its polarization, orientation, and the degree of focusing given by $\rho_0 = R_0 / w_0$ where $w_0$ is the beam waist radius. Varying these parameters one observes either a single spot or a ring pattern.  

It is worth noting that the ring intensity pattern combined with a nontrivial polarization structure of the conically refracted light has been already employed in trapping of micro-objects. However, in this case the beam was used to form traditional  tweezers  for two-dimensional manipulation  of particles~\cite{Dwyer:opex:10,Dwyer:opex:12}. Here we are interested in the three-dimensional 3D structure of the conically refracted light.  In the paraxial regime the light intensity distribution at a distance $z$ behind the biaxial crystal can be analytically described as~\cite{Belskii:os:78}
\begin{equation}\label{CP}
I_{CP}(\rho,z)= |B_C|^2+|B_S|^2
\end{equation}
for a circularly  polarized beam,  and
\begin{equation}\label{LP}
I_{LP}(\rho,z)= I_{CP}+(B_CB_S^*+B_C^*B_S)\cos(2\Phi-(\varphi+\varphi_C))
\end{equation}
for a linearly polarized input beam.
Here $\vec{\rho}=(\cos\varphi, \sin\varphi)$ represents  transverse coordinates, $\Phi$ is the angle of the initial linear polarization and  $\varphi_C$ defines the orientation of the plane of the crystal's axes. Functions $B_S(\rho,z)$ and $B_C(\rho,z)$ are defined as
\begin{eqnarray}
B_C(\rho,z)& =& \int_0^{\infty}\eta a(\eta)\mbox{e}^{-i\frac{z}{4}}\eta^2\cos(\eta\rho)J_0(\eta\rho_0)d\eta \\
B_S(\rho,z)&=& \int_0^{\infty}\eta a(\eta)\mbox{e}^{-i\frac{z}{4}}\eta^2\sin(\eta\rho)J_1(\eta\rho_0)d\eta \\
a(\eta)&=&\int_0^{\infty}\rho E^{in}(\rho)J_0(\eta\rho)d\rho
\end{eqnarray}
where $E^{in}(\rho)$ is the amplitude of the input beam and  $\eta=|\vec{k}_\perp|w_0$.
\begin{figure}
\centerline{\includegraphics[width=8cm]{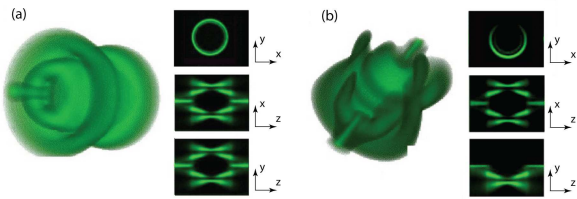}}
 \caption{\label{setup}(color online) Calculated light intensity distribution  of conically refracted light beam in biaxial crystal. (a) circular  polarization (Eq.(\ref{CP}) leads to a fully closed optical bottle; (b) linearly polarized input beam (Eq.\ref{LP}) leads to an open-top optical bottle. The insets depict the principal  cross sections of the light intensity distribution.  Here $\rho_0=$. }
\end{figure}
According to Eq.(\ref{CP}), for a circularly polarized input beam the light intensity distribution behind the crystal forms a cylindrically symmetric structure. One can show~\cite{Turpin:13} that for $\rho_0 \gg 1$, the previous equations describe the formation of the well resolved ring at the focal plane as well as the bottle beam between the two Raman spots. This situation is  illustrated in Fig.1 (a)  where we show the 3D light  intensity distribution behind the crystal for $\rho_0=10$. One can clearly observe the existence of a dark region surrounded by high light intensity.

A more complex intensity pattern can be realized with a linearly polarized input beam. From the Eq.(\ref{LP}) one finds that the linear polarization  results in the loss of perfect cylindrical symmetry. Instead of rings the transverse light intensity distribution has a form of crescents over certain finite distance. From the 3D perspective this indicates the formation of a hole in the side of the otherwise perfect bottle beam.  This situation is depicted in Fig.1(b). The gap in the otherwise homogeneous intensity pattern is clearly visible. What is more important, the angular position of the gap can be varied  by rotating the azimuth of the linear  polarization of the input beam. The appearance  of the hole in  an optical bottle and the ability to vary its position provides a unique opportunity to control the trapping conditions. Opening  the hole in  the bottle  would allow one to quickly  load particles into the trap. The trap can then be closed by switching over to circular  polarization. Reverting to the linear polarization would enable opening the hole in the trap at a desired angular location to  unload the particles.

\section{Experiment}

We tested experimentally the practical  suitability  of the above described optical bottle beams for trapping of    airborne microscopic particles. The experimental setup is schematically depicted in Fig.2. The light beam from a cw laser ($\lambda$=532nm, input power 100mW) passes  through a $\lambda/2$ and $\lambda/4$ wave plates and then, after focusing with 100~\rm{cm} positive lens, propagates  along the optical axis of a monoclinic KTP crystal ($l=10$mm and $\alpha=0.010$~\rm{mrad}) cut perpendicular to one of its  optic axes, giving $\rho_0 \approx 12$.
Light emerging from the crystal is imaged with a CCD camera. We start with a circularly polarized  input beam to create a perfect, cylindrically symmetric optical bottle. In order to visualize the optical  bottle the camera  was translated axially with a 10$\mu$m step and at each step the transverse light intensity distribution is recorded and stored in the computer. A sequence of 75 intensity slices is then used to reconstruct   the full 3D  structure of the bottle. The result is depicted in Fig.3(a) and accompanying movie. It is evident that the conically refracted  light does form an optical bottle with a dark central region entirely surrounded by light. The  transverse  size and the length of the bottle could be adjusted  by varying the collimating optics as well as the position of the crystal.
 \begin{figure}
\centerline{\includegraphics[width=8cm]{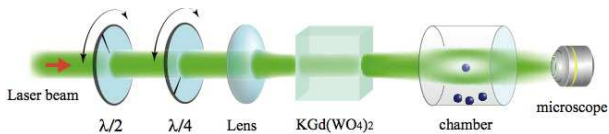}}
 \caption{\label{setup}(color online) Experimental setup. }
\end{figure}
 In the next step the polarization of the input beam is changed to linear. The corresponding 3D light intensity distribution is shown in Fig.3(b) and accompanying movie. The  structure is no longer cylindrically symmetric, with the top wall of the bottle featuring  an opening, in agreement with the theoretical prediction  [see Fig.1(b)].

\begin{figure}
\centerline{\includegraphics[width=8cm]{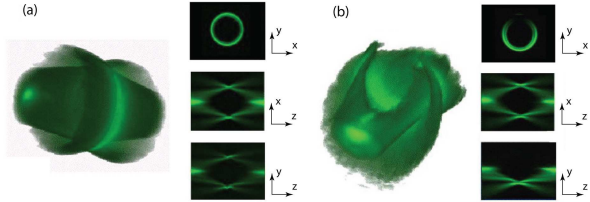}}
 \caption{\label{setup}(color online) Experimentally recorded  3D structure of an optical bottle beam formed by the conical refraction of light in a KGd(WO$_4$)$_2$ crystal. (a) The bottle is fully closed for a circularly polarized incident beam and (b) open for a linearly polarized incident beam. The hole  in the top wall of the ``bottle'' is clearly visible. The insets depict the principal  cross sections of the light intensity distribution.  Here $\rho_0 \approx 12$}
\end{figure}
 We used  the optical bottle depicted in Fig.3 to demonstrate trapping and manipulation of airborne light absorbing particles.  Such particles can be efficiently confined by employing the photophoretic force~\cite{Jovanovic:jqsr:09,Shvedov:opex:11}. In this case the  illumination of particles leads to their heating and nonuniform temperature distribution. Interaction with the surrounding air results in appearance of the photophoretic force which tends to repel particles from the high intensity region. We have recently demonstrated efficient photophoretic trapping and transport of micrometer size particles over large distance of tens of centimeters~\cite{Shvedov:prl:11,Shvedov:apl:12}.  In our experiments with the optical bottle we used glass shells covered with a thin layer of carbon (~200nm) in order to enhance the  light absorption. The external diameter of the shells  varied  ranging from a few to tens of micrometers.  To prevent accidental air flow  from affecting the trapping the optical bottle was formed inside a transparent glass cell placed  immediately behind the biaxial crystal. The axially located CCD camera recorded images of the particles inside the optical bottle. In order to speed up the trapping process the spheres were made floating in the air.

 \begin{figure}
\centerline{\includegraphics[width=8cm]{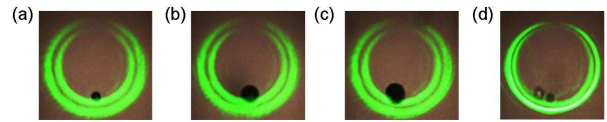}}
 \caption{\label{setup}(color online) Experimental images of hollow glass shells of different size  trapped inside the bottle beam. In (a-c) shell diameter is 21~\rm{$\mu m$}, 45~\rm{$\mu m$}, and 50~\rm{$\mu m$}, respectively. (d) Simultaneous trapping of two glass spheres. Here $\rho_0 \approx 12$.}
\end{figure}

We found that while a particle could be trapped using either a fully closed (circular polarization) or open (linear polarization) bottle, the loading process was much quicker in the latter case. As the internal diameter of the bottle was rather large (200~$\mu m$) the bottle could accommodate a great variety of trapped spheres. In Fig.4 we show three examples of particles with different size  confined in the trap. Because of gravity they are all located at the bottom of the bottle.  The trapping was generally very robust, with the particles stably resting on the lower ``wall''. However, we found that sometimes the trapped particles oscillated inside the trap  with the oscillation frequency increasing  with the trapping power. Such a dynamics was observed in the case of trapping complex objects such as those formed by two connected glass spheres.
\begin{figure}
\centerline{\includegraphics[width=8cm]{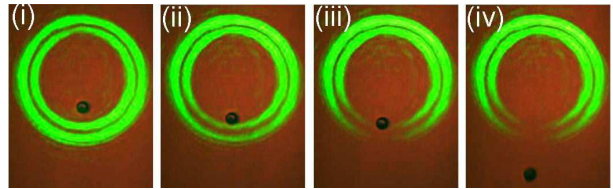}}
 \caption{\label{setup}(color online) Experimentally recorded  sequence of images illustrating  unloading the trapped glass sphere from the bottle. The sequence of graphs (i)-(iv) represents different stages of the opening the gap in the bottom side of the bottle.  Here $\rho_0 \approx 12$. }
\end{figure}

The ability to open or close the bottle at will by varying the polarization of input beam gives a unique opportunity to use the opening in the bottle not only to easily trap micro-object but also unload the trap. Such functionality is demonstrated in Fig.5.  The image sequence  represents various stages of closing the hole  in the upper wall of the bottle while simultaneously  opening it in its bottom wall. It is clearly seen that the initially stably trapped sphere drops out of the trap under gravity when the hole in the bottom wall opens.

\section{Conclusions}
In summary, we have employed  conical refraction of light in a biaxial crystal to create an optical bottle for  trapping and manipulation of airborne  particles. We have demonstrated that by just varying the polarization of the input light the  bottle can be  opened and closed. We have also  experimentally demonstrated stable photophoretic trapping and controllable loading  and unloading of particles in the trap.

\section{Acknowledgement}
This work was supported by the Australian Research Council, and the Spanish MICINN and the Catalan Government through the contracts FIS2011-23719 and SGR2009-00347, respectively. A. T. acknowledges financial support through grant AP2010-2310 from the MICINN.

\end{document}